\documentclass[twocolumn]{aastex631}

\usepackage{graphicx}
\usepackage{eurosym}
\usepackage{natbib}
\usepackage{booktabs}
\usepackage{amsmath}
\usepackage{graphics,graphicx}
\usepackage{hyperref}
\hypersetup{urlcolor=blue}
\usepackage{cleveref}
\usepackage{eurosym}
\usepackage{booktabs}
\usepackage{wrapfig}
\usepackage{tabularx}

\newcommand{\rsun}{$R_\odot$}
\newcommand\be{\begin{equation}}
\newcommand\ee{\end{equation}}

\def\PP{{\cal P}}

\def\dd{{\rm d}}

\def\PP{{\cal P}}

\shorttitle{Time Variation of the Tachocline}
\shortauthors{Basu et al.}

\submitjournal{ApJ}

\begin{document}

\title{Time Variation of the Solar Tachocline}

\author[0000-0002-6163-3472]{Sarbani Basu}
\affiliation{Department of Astronomy, Yale University, PO Box 208101, New Haven, CT 06520-8101, USA}

\author[0009-0006-6564-7798]{Wesley Antônio Machado Andrade de Aguiar}
\affiliation{Department of Computer Science, Yale University, PO Box 208285, New Haven, CT 06520-8285, USA}

\author[0000-0003-1531-1541]{Sylvain G. Korzennik}
\affiliation{Center for Astrophysics $|$ Harvard \& Smithsonian, Cambridge, MA 02138, USA}

\begin{abstract}

We have used solar oscillation frequencies and frequency splittings  obtained over solar cycles~23, 24 and the rising phase of solar cycle~25 to investigate whether the tachocline properties (jump {i.e., the change in the rotation rate across the tachocline}, width and position) show any time variation. We confirm that the change in rotation rate across the tachocline changes substantially, however, the change does not show a simple correlation with solar cycle unlike, for instance, changes in mode frequencies. The change during the ascending phase of solar cycle~25 is almost a mirror image of the change during the descending part of solar cycle~24, tempting us to speculate that the tachocline has a much longer period than either the sunspot or the magnetic cycle.  We also find that the position of the tachocline, defined as the mid-point of the change in rotation rate, showed significant changes during solar cycle~24. The width of the tachocline, on the other hand, has showed significant changes during solar cycle~23, but not later. The change in the tachocline becomes more visible if we look at the upper and lower extents of the tachocline, defined as (position~$\pm$~width). We find that for epochs around solar maxima and minima, the extent decreases before increasing again --- a few more years of data should clarify this trend. Our results reinforce the need to continue helioseismic monitoring of the Sun to understand solar activity and its evolution.

\end{abstract}


\section{Introduction}\label{sec:intro}

Helioseismic analyses have shown that the convection zone (henceforth CZ) of the Sun rotates differentially, while the radiative interior rotates like a solid body \citep[etc.]{schou}. Connecting the two zones is a thin shear layer, known as the tachocline.  

Neither the origin of the tachocline nor the role it plays is well understood. This layer is often cited as a key region in the process of magnetic field generation in the Sun and other solar-like stars, as it is a region of strong shear and therefore capable of converting weak poloidal fields into strong toroidal fields. Some dynamo models of the Sun \citep{dikpati1999, chatterjee2004, guerrero2008} assume that the large-scale toroidal magnetic field generated at the tachocline can, when strong enough, move up to the surface and manifest as sunspots. 
In fact, in 3D MHD simulations of the solar dynamo that include the tachocline, most of the magnetic field develops at the base of the convection zone \citep{guerrero}. It has also been argued that the tachocline plays a key role in establishing the period of the solar cycle, the origin of torsional oscillations, and the scaling law of stellar magnetic fields as a function of the Rossby number \citep{guerrero2017}.
 
If indeed the tachocline is responsible for the changing magnetic fields in the Sun, one could expect that the tachocline varies on solar-cycle timescales. It has been shown by \citet{basuantia2019} that the change in the rotation rate between the solar CZ and the radiative zone (the ``jump'') is a function of time; however, the temporal dependence is not a simple correlation with magnetic activity. The jump followed different trends during solar cycle~23 than solar cycle~24 and, furthermore, had different values at the same level of solar activity. 
\citet{basuantia2019} could not, however, detect any changes in the position of the tachocline or the width of the tachocline; the uncertainty of their result was too large due to the lack of precision of the data sets they used. 
The data sets used in that study were splitting coefficients obtained with 108-day long time series from the Global Oscillation Network Group \citep[GONG:][]{gong} along with those obtained using 72-day long time series from the Michelson Doppler Imager \citep[MDI:][]{mdi} on board the Solar and Heliospheric Observatory (SOHO) and the Helioseismic and Magnetic Imager \citep[HMI:][]{hmi} on board the Solar Dynamics Observatory (SDO). An earlier attempt by \citet{antiabasu2011} using data from the same sources was also unsuccessful in detecting any clear changes in the thickness of the tachocline, although they did observe changes in the jump and possible changes in the position.

Changes in the thickness of the tachocline could help distinguish between dynamo models of the Sun. For instance, \citet{dikpati2001} claim that if toroidal fields are concentrated in relatively narrow bands that migrate towards the equator as the sunspot cycle advances, then they should be accompanied by a ``thickness front'' advancing at the same rate. The position of the tachocline may also change --- among the different scenarios proposed to explain the relatively weak solar activity between 2005 and 2010, one hypothesis attributes it to  small ``pulsations'' in the position of the tachocline \citep{dejager}.

The time series used in earlier investigations into tachocline properties were too short to obtain frequency splitting coefficients precise enough to reliably determine changes in the position and thickness of the tachocline. The situation has since changed and  splitting coefficients derived from longer time series are now available, and in addition we have data spanning more than two solar cycles. In this paper, we use frequency splittings obtained with different lengths of helioseismic time series to determine temporal variations in the position and thickness of the tachocline.

The rest of this paper is organized as follows. We describe the data used in \S~\ref{sec:data}, the tachocline model, and how we determined the tachocline properties in \S~\ref{sec:model}. Our results are described in \S~\ref{sec:res}, and we discuss the results in \S~\ref{sec:concl}.

\section{Data Used} \label{sec:data}

For this work, we use solar oscillation frequencies derived from observations obtained with the ground-based GONG network and the space-borne MDI and HMI instruments on board the Solar \& Heliospheric Observatory (SOHO) and the Solar Dynamics Observatory (SDO) respectively. 
MDI observations started in 1995 and ended in 2010, hence they cover solar cycle 23, 
while HMI observations started in 2010 and are continuing, hence HMI observations cover cycles 24 and 25. 
There is a one-year overlap between MDI and HMI observations. 

We use rotational frequency splittings obtained by an independent data reduction pipeline \citep{syl1, syl2, syl3, syl4, syl5, sgk2018, sgk2023}, which we refer to as the ``SGK'' pipeline. The SGK pipeline derives mode parameters from time series that are multiples of 72 days, therefore we use mode frequencies and splittings obtained for time series lengths of $32 \times,\; 16 \times,\; 10\times,\; 8 \times$ and $5 \times 72$ days, or  approximately 6, 3, 2, 1.5 and 1 year, respectively.  The choice of using multiples of 72-days was dictated by the standard time series length adopted initially by the MDI team, and later adopted by the HMI team. At the time, a time series of 72 days was considered to be long enough to produce a power spectrum with high signal-to-noise ratio, but short enough to provide ``good'' temporal resolution of the inferred frequencies and frequency splittings. The MDI and HMI teams later started analyzing 360-day (i.e., $5\times 72$-day) long time series too. We also use splittings from the respective teams pipelines (henceforth referred to simply as ``pipeline data'') to compare to results obtained with the splittings inferred with the SGK pipeline.

The SGK pipeline predominantly derives mode parameters by fitting asymmetric profiles to peaks in the power spectra, though some data sets have also been fitted  using symmetric profiles. The MDI and HMI projects' pipeline, on the other hand, generally fits symmetric profiles, though some 360-day long sets have been fitted with asymmetric profiles.

{The starting dates of all SGK sets are equally spaced in time, aligned with the first 72-day long MDI set (i.e., 1995.05.01). The MDI and HMI $5\times72$-day long sets from the projects' pipeline, however, do not always have equal spacings, and some of the start dates of the project's and SGK sets differ. The start times of all sets that we analyzed are listed in the tables in Appendix~\ref{app:tabs}. }

The data sets are available in the usual form, where the  mode frequencies and their splittings are expressed as follows:
\begin{equation}
\nu_{nlm}
= \nu_{nl} + \sum_{j=1}^{j_{\rm max}} a_j (n,l) \, \PP_j^{(l)}(m), 
\label{eq:eq_split}
\end{equation}
where $\nu_{nl}$, or the central frequency of a mode of degree $l$ and radial order $n$, is determined by the spherically symmetric part of solar structure, $a_j$ are splitting coefficients and
$\PP_j$ are re-scaled Clebsch-Gordon coefficients \citep[see][]{SchouEtal:2002}. In this decomposition, the odd-order $a_j$ are caused  by the solar rotation, while the even-order coefficients contain the signature of asphericity and magnetic fields. In order to compare with earlier work, we convert the $a$ coefficients to $c$ coefficients as defined by \citet{ritz}.

\begin{figure}[]
    \centering
    \includegraphics[width=3.3 true in]{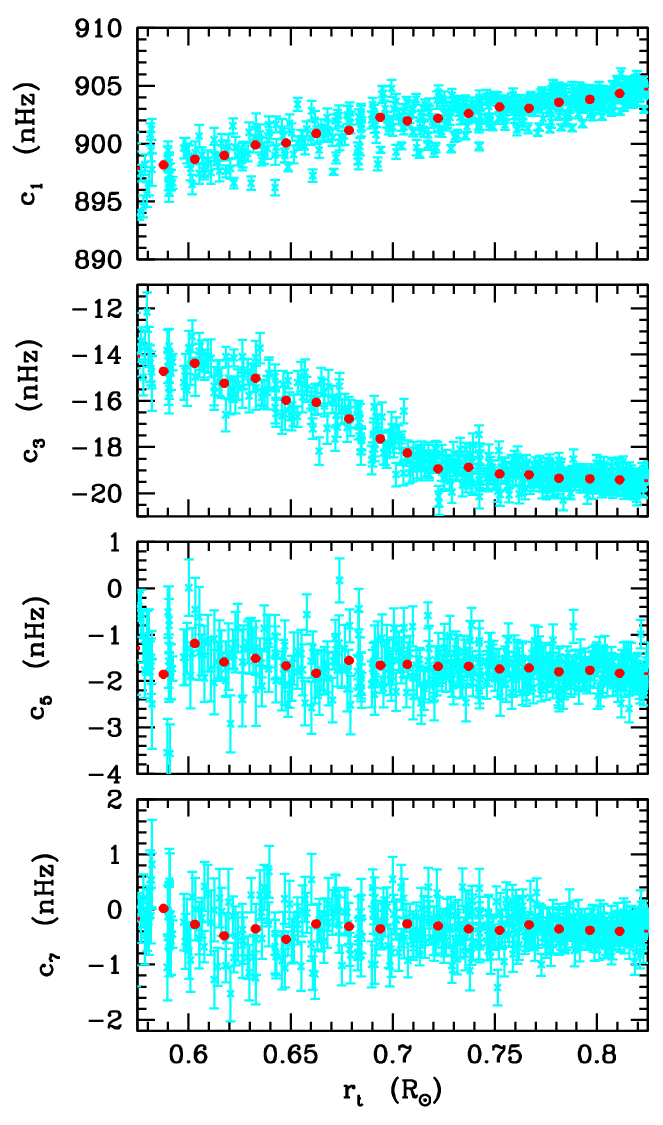}
    \caption{Odd-order splitting coefficients, $c_1$--$c_5$, derived from a $16\times 72$-day long time series plotted as a function of the lower turning points of the modes. The {cyan} points are the individual values, while the {red points show values averaged over $r_t$ bins of width 0.015\rsun. The error bars on the red points are not visible because they are similar to the size of the symbols.} }
    \label{fig:split}
\end{figure}

The first four odd-order splitting coefficients for a $16\times 72$-day long GONG time series plotted as a function of the lower-turning point\footnote{{The lower turning point, $r_t$, is the deepest point that an acoustic mode propagates to. It can be evaluated through the relation $ c_s^2/r_t^2=\omega^2/L$, where $c_s$ is the speed of sound, $\omega\equiv 2\pi\nu$, and $L=l(l+1)$.}}
are shown in Fig.~\ref{fig:split}. That figure shows that the clearest signature of the tachocline is in the $c_3$ splittings, and is essentially the same as that shown in Fig.~10 of \citet{basu1997}. There appears to be a small signature in $c_1$, while the other splitting coefficients are too noisy to show such a signal. Consequently, we only use the $c_3$ coefficient in the work presented here. Furthermore, we only use modes with frequencies between 1.5 mHz and 3.5 mHz that have lower turning points between 0.55\rsun\ and 0.85\rsun\ for the range of degrees that is covered by the data set. 
This subset of mode-splitting coefficients give good coverage of the tachocline, while keeping the  uncertainties low; this also removes the need to account for the near-surface shear layer in the tachocline model.

\section{Tachocline model and analysis technique} \label{sec:model}

We perform a forward analysis to determine tachocline properties. 
Following \citet{ritz}, we express the solar rotation velocity, \citep[see also][]{abc}, as
\begin{equation}
v_{\rm rot}(r, \vartheta)=\Omega(r,\vartheta)r \sin\vartheta =
-\sum_{s=0}^\infty w_{2s+1}(r)\frac{\partial}{\partial \vartheta} Y^0_{2s+1},
\label{eq:rot}
\end{equation}
where $r$ is the radius, $\vartheta$ the co-latitude, $Y^m_l(\vartheta)$ are spherical harmonics and $w_s(r)$ are functions that are related to the splitting coefficients as follows:
\begin{equation}
    c_s^{(n,l)}=\int_0^R w_s(r)K_s^{(n,l)}r^2 \dd r,
    \label{eq:ccoef}
\end{equation}
where the kernels, $K_s^{(n,l)}$, are known functions of the displacement eigenfunctions of the modes \citep[see Eq.~5 of][]{abc}.

Eq.~\ref{eq:rot} essentially separates the radial and angular dependence of the rotation rate, and Eq.~\ref{eq:ccoef} shows how $c_3$ can can be calculated given a one dimensional rotation rate $\Omega(r)$, which in our case, is the model of the tachocline. At any colatitude $\vartheta$, the rotation rate
$\Omega(r,\vartheta)$ is simply 
\begin{equation}
    \Omega(r,\vartheta)=-\Omega(r)\frac{\partial}{\partial\vartheta}Y^0_3=
    \Omega(r)\frac{3}{4}\sqrt{\frac{7}{\pi}} (1-5\cos^2\vartheta),
    \label{eq:lat}
\end{equation}
implying that there is no contribution from $c_3$ at a latitude of approximately 26.56$^\circ$.
Following \citet{basu1997}, we model the tachocline as a sigmoid and parametrize the rotation rate around the tachocline as
\begin{equation}
    \Omega(r)=\frac{\delta\Omega}{1+\exp[{-(r_d-r)/w_d}]},
    \label{eq:tach}
\end{equation}
where $\delta\Omega$ is the jump in the rotation rate between the convection zone and the interior, $r_d$ is the position of the tachocline, defined as the mid-point of the transition (or discontinuity), and $w_d$ is the width of the transition layer.
With this parametrization, the rotation rate changes from 0.269 to 0.731 (i.e., from $1/(1+e)$ to $1-1/(1+e)$) of $\delta\Omega$ between radii of $r=e_d-w_d$ and $r=r_d+w_d$. 
{The tachocline is located approximately at the base of the convection zone, i.e., $r_d$ is about $0.71$\rsun.}
Note that \citet{abc}, \citet{antiabasu2011}, and \citet{basuantia2019} used essentially the same model of the tachocline, but had co-latitude dependent terms since they did a full 2D fit. 
Also note that in this parametrization of the tachocline, the definition of the width is different from that used by \citet{agk1996} and \citet{paulchar}, where the tachocline was parametrized by an error-function, ${\rm erf}(r)$, namely 
$\Omega(r)={\delta\Omega}( 1+ {\rm erf}(2(r-r_d)/{w_d})/2$, where the width corresponds to the region where the rotation rate changes from 0.08 to 0.92 of the jump, $\delta\Omega$. For small widths, $w_d$, one can show that $w_d^{\rm erf}\approx 5\,w_d^{\rm sigmoid}$. 

We used two minimization techniques to estimate the three parameters of the tachocline, namely $\delta\Omega$, $r_d$ and $w_d$. One is based on a search within a pre-computed grid of models, while the another uses  a simulated annealing (SA) optimization. These two methods are described in Appendix~\ref{app:fitting}. The grid-based method is very fast, while the SA method, even with only three parameters, is slow. However, as shown in Appendix~\ref{app:fitting}, the grid-based method does not recover the width of the tachocline very well, and since determining possible changes in the width of the tachocline is a major goal of this work,  we only report results obtained with  the SA method. 

\begin{figure*}[h]
    \centering
    \includegraphics[width=0.9\textwidth]{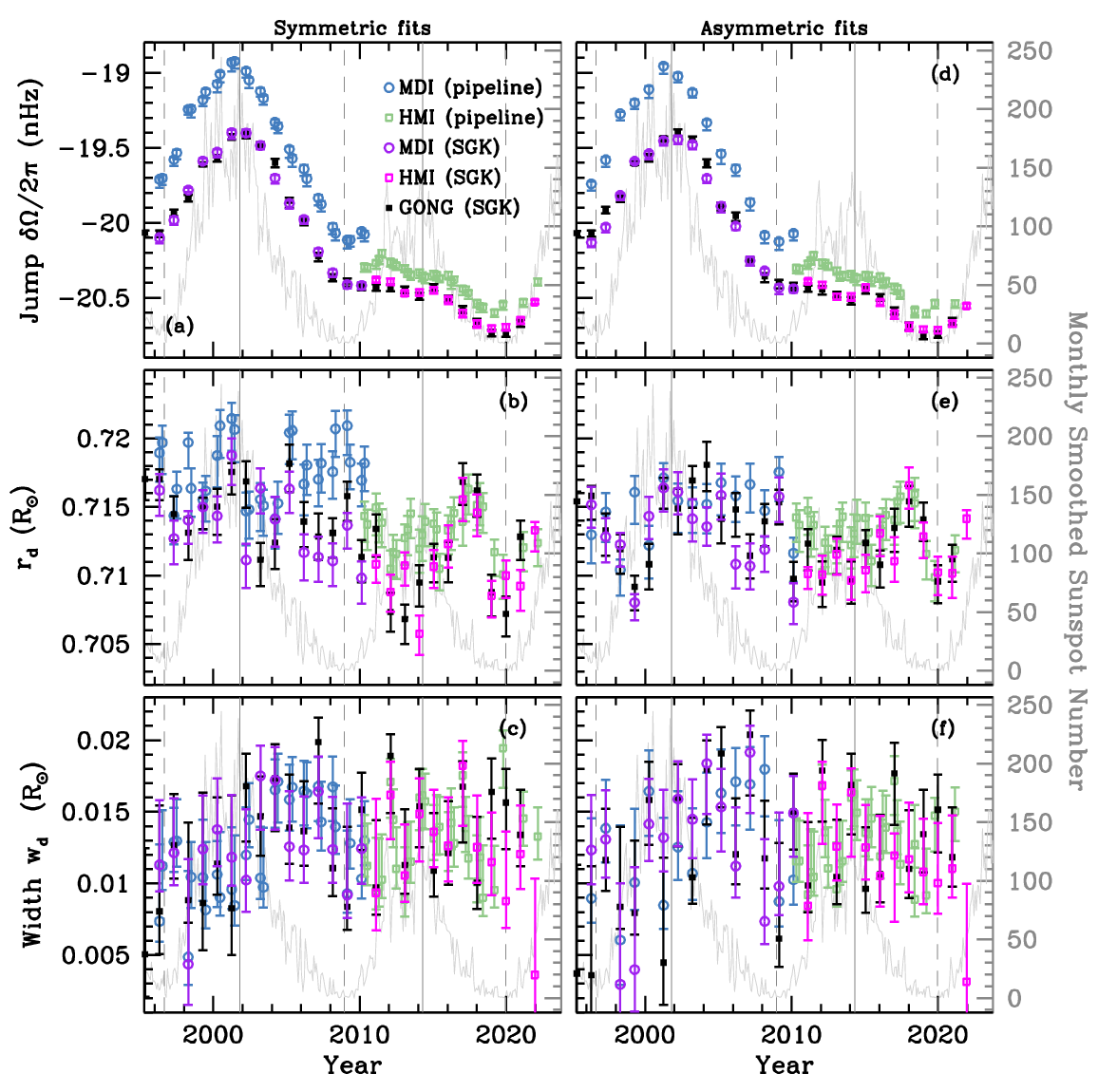}
    \caption{Tachocline properties estimated using data sets obtained with $5\times72$-day long time series plotted as a function of time, using the time at the  mid-point of the observational window. The panels on the left were obtained for data with symmetric fits to the mode profile, while those on the right with asymmetric profiles. 
    The different colors and symbols correspond to data sets corresponding to GONG, MDI or HMI observations using either the MDI or HMI projects' pipelines or the SGK pipeline as shown in the legend. {The gray curve shows the monthly smoothed sunspot numbers from  the Royal Observatory of Belgium, Brussels, whose scale can be read on the right-hand axes.}
    The vertical solid lines mark the times of solar maximum, while the vertical dashed lines mark the times of solar minimum determined from smoothed values of activity proxies.}
    \label{fig:05e}
\end{figure*}

\section{Results} \label{sec:res}

Since we are interested in examining the temporal variation of the tachocline, we examine results obtained with data sets corresponding to the shortest time series that we are considering, i.e., the $5\times72$-day long ones, before examining the longer sets.

\subsection{Results from $5\times72$-day sets}\label{subsec:short}

The 360-day-long data sets are useful because they have been processed both with the MDI and HMI project pipelines, as well as the SGK pipeline. Furthermore, the mode parameters were obtained using both symmetric and asymmetric profiles. The resulting tachocline parameters are presented in Fig.~\ref{fig:05e} and tabulated in Tables~\ref{tab:05a} and \ref{tab:05b} in Appendix~\ref{app:tabs}.

One can see from Fig.~\ref{fig:05e} that $\delta\Omega$, the jump across the tachocline, is the most well-determined property, and that the jump shows a very clear temporal variation, although the changes during cycle~23 were quite different from those during cycle~24. This is consistent with what was found by \citet{basuantia2019}. We also find a systematic difference between the jump,  obtained with the project pipeline data sets and those obtain with the SGK pipeline. \citet{basuantia2019} had seen a similar systematic difference between results from the GONG, MDI and HMI pipelines' data sets. Since we do not see any such difference between GONG, MDI and HMI observations processed with the SGK pipeline, we are led to believe that the root of this difference lies in the choices made in the projects' pipelines. Furthermore, one can also see that there is a systematic difference between the MDI and HMI pipeline results during the time when both projects were observing. While such systematic differences add to the uncertainty of $\delta\Omega$, we can see that it does not affect inferences about the temporal variation. We also see that results obtained when using splitting coefficients obtained using symmetric and asymmetric mode profiles are consistent with each other. 

\begin{figure}
    \centering
    \includegraphics[width=3.5 true in]{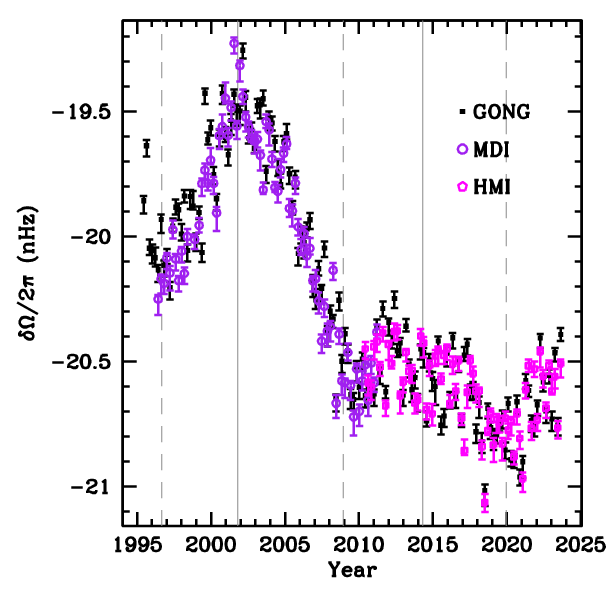}
    \caption{The jump across the tachocline derived from the $72$-day long time series data sets using GONG, MDI or HMI observations and the SGK pipeline (black dots, purple circles and magenta circles respectively). The solid and dashed vertical lines mark respectively the positions of solar maxima and solar minima.}
    \label{fig:01e}
\end{figure}

We also have a few data sets corresponding to the rising part of cycle~25, and we see that the change in the jump is again trending  differently from the previous solar cycles.
Indeed, the absolute value of the jump is decreasing, and has almost overcome the increase seen during the falling part of cycle~24. It will be interesting to see if by the end of cycle~25 the value of this jump will be back to the value it had at the end of cycle~23. If so, it would indicate the presence of a much longer activity cycle at the tachocline. Since shorter time series data allow us to add to the cycle~25 timeline, we show the jump results obtained with $72$-day long time series in Fig.~\ref{fig:01e}; although these results are noisier, the upward trend is clearly visible. From that same figure, one is tempted to infer that results obtained with GONG observations acquired during the last year of cycle~22 show a fall in $\delta\Omega$, which then becomes a rise in the ascending phase of cycle~23. These values are presented in Tables~\ref{tab:01e}, \ref{tab:mdi01e} and \ref{tab:hmi01e} in Appendix~\ref{app:tabs}.

\begin{figure}[t]
    \centering
    \includegraphics[width=3.5 true in]{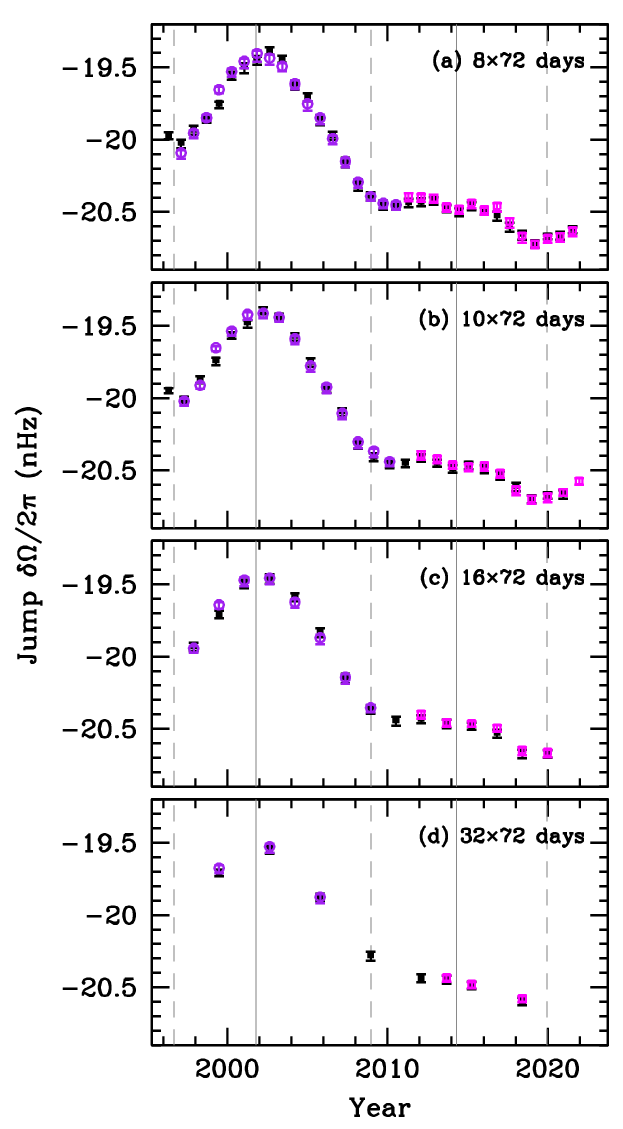}
    \caption{The jump in the rotation rate across the tachocline for data sets corresponding to the approximately 1.5-year, 2-year, 3-year and 6-year long time series data sets for GONG, MDI or HMI and the SGK pipeline. Like in Fig.~\ref{fig:01e}, the black dots correspond to the GONG results, purple circles to MDI results and magenta circles to HMI results. The vertical solid and dashes lines mark epochs of solar maxima and minima, respectively.}
    \label{fig:alljump}
\end{figure}

\begin{figure*}[t]
    \centerline{
    \includegraphics[width=3.5 true in]{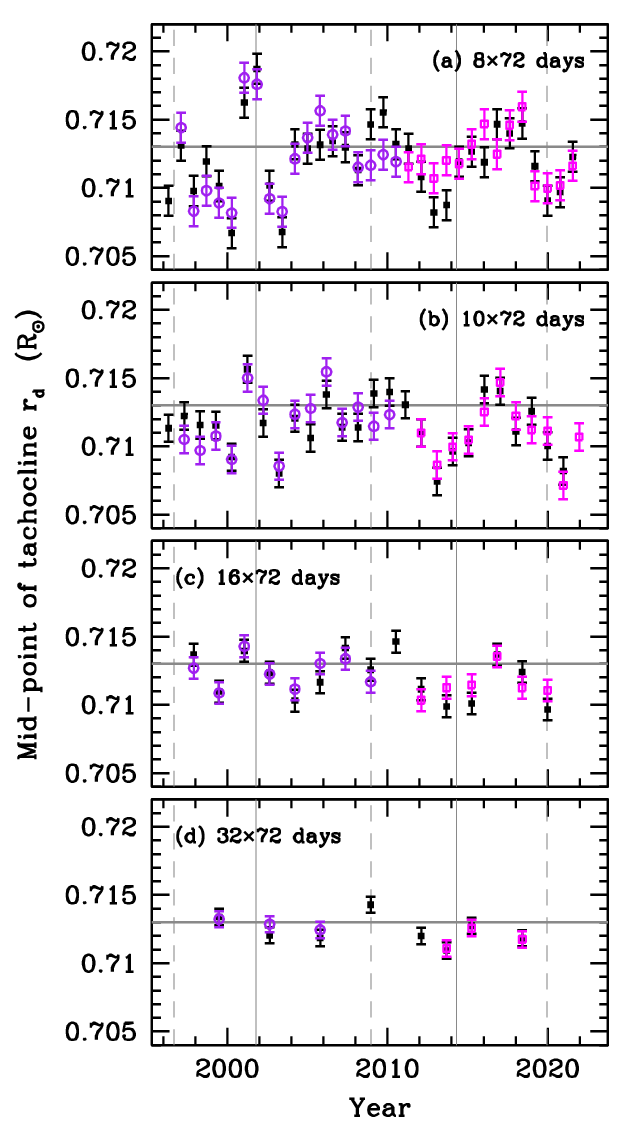}
    \includegraphics[width=3.5 true in]{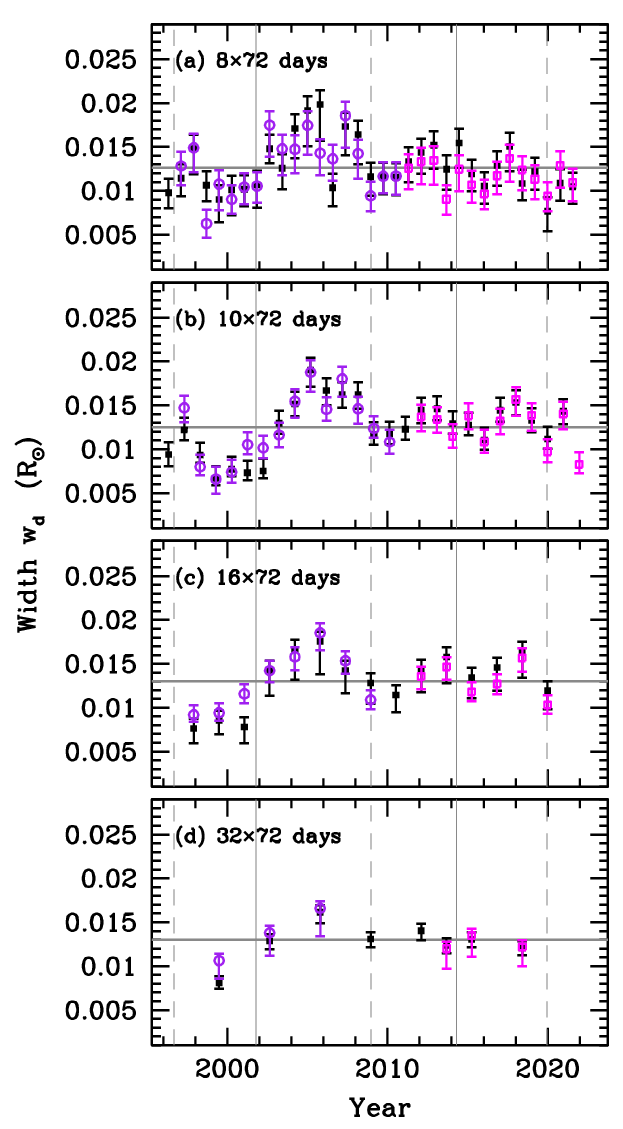}}
    \caption{The position (left panels) and the width (right panels) of the tachocline plotted as a function of time for data sets of different time series length, i.e., $8\times, 10\times, 16\times$ and $32\times72$-day (top to bottom) and the SGK pipeline. Like in Fig.~\ref{fig:01e}, the black dots correspond to the GONG results, purple circles to MDI results and magenta circles to HMI results.
    The vertical solid and dashes lines mark epochs of solar maxima and minima, respectively. The horizontal line in the panels on the left mark the position of the CZ base. The horizontal lines in the panels on the right are an average of all the points in each panel.}
    \label{fig:all}
\end{figure*}

Returning to the $5\times72$-day long results, we find that the derived estimates of the tachocline mid-point are quite noisy, making it difficult to state whether the temporal variations are statistically significant over the entire timeline. The results derived from symmetric mode-profile fits appear to be more scattered than those derived from asymmetric mode-profile fits. Whether the derived positions during cycle~25 are near mirror-image of the values during the declining phase of cycle~24, like the jump, is more difficult to discern, though it appears to be so. As noticed before by \citep{nssl, asph}, the results derived from GONG and HMI observations agree better than those derived from GONG and MDI observations. 

The scatter in the results for the width of the tachocline, $w_d$, is also quite large. Consequently, while it appears that the temporal trend during cycle~23 was different, we cannot ascertain whether they are statistically significant. However, we do note that the official pipeline and SGK results are in quite good agreement.

\subsection{Results from longer time series data}

Since the data derived from the $\approx 1$-year time series are quite noisy, we determined the tachocline parameters from longer time series data,  although, we lose temporal resolution.

Even though the jump results are clearly significant even for results obtained with short time series data sets, for the sake of completeness, we show the result for jump using longer time series data sets in Fig.~\ref{fig:alljump}. These results are also presented in Tables~\ref{tab:08}, \ref{tab:10}, \ref{tab:16} and \ref{tab:32}. The general temporal variability is the same for all the data sets, but the change becomes smoother as the length of the analyzed time series increases. {The longer time series decreases the temporal resolution of the data. This causes time variations on scales shorter than the length of the time series to be averaged out. This averaging also decreases the amplitude of the variations} 

The estimated positions and widths of the tachocline are shown as a function of time in Fig.~\ref{fig:all} for the different instruments. 
There was a statistically significant change in the position of the mid-point of the tachocline in solar cycle~24, which seems to be continuing in solar cycle~25. The issue is more confusing when looking at cycle~23. Indeed, the $8\times72$-day and the $10\times72$-day long results show a sudden change sometime before 2002. Since this change is seen when using both GONG and MDI data sets, it is tempting to believe that this change is real and not a result of the fitting technique either, as can be seen from Fig.~\ref{fig:comp}(b). Such a change is also seen in results derived from the $5\times72$-day long data sets (see Fig.~\ref{fig:05e}), but those have much larger uncertainties, and hence the sudden increase no longer stands out. Of course, this feature is smoothed out when longer time series data sets are used, thus, we cannot say with any certainty whether there was a sudden change in the tachocline position around 2001-2002. 

As hinted by results obtained with the data sets corresponding to the  $5\times72$-day long time series, there is a time variation in the width of the tachocline which is best seen when using the $10\times72$-day long time series data sets --- results derived from shorter time series are noisy, while results corresponding to longer ones tend to average out a lot of the variation. As can be seen from Fig.~\ref{fig:all}, the variations of the width during cycle~23 were very different from the variations during cycle~24 which are not statistically significant. 

\section{Discussion} \label{sec:concl}

We have determined tachocline parameters from helioseismic mode splittings obtained with various lengths of time series. This allows us to beat down the noise in the analysis in order to examine whether the position of the mid-point of the tachocline and its thickness vary with time. 

\citet{basuantia2019} has already shown that the jump in the rotation rate changes with time. This work confirms that, and extends the results to the ascending phase of solar cycle~25. The change during cycle~25 seems to be a mirror image of the change during the descending phase of cycle~24. This leads us to speculate that there could be a longer magnetic cycle present at the tachocline. We need to wait for additional data acquired beyond cycle-25, namely during the ascending part of cycle~26, to confirm this.

\begin{figure}[t]
    \centering
    \includegraphics[width=3.25 true in]{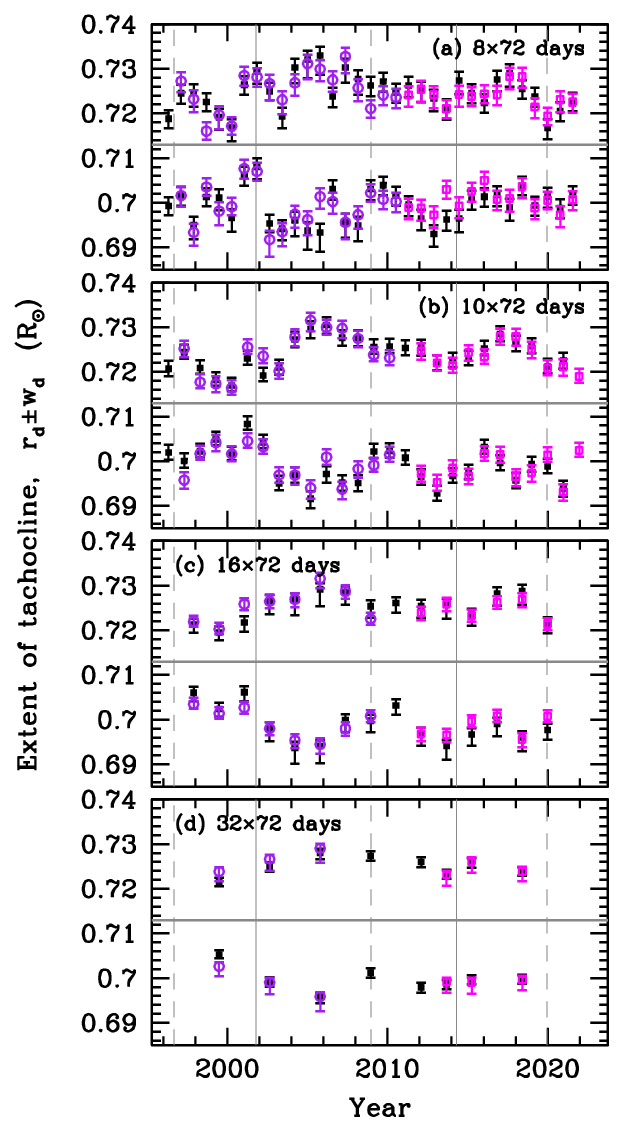}
    \caption{ The upper and lower boundaries of the tachocline, i.e., $r_d+w_d$ and $r_d-w_d$, plotted as a function of time for data sets obtained with different time series lengths, i.e., $8\times, 10\times, 16\times$ and $32\times72$-day (top to bottom) and the SGK pipeline.
    Like in Fig.~\ref{fig:01e}, the black dots correspond to the GONG results, purple circles to MDI results and magenta circles to HMI results.
    The vertical solid and dashes lines mark epochs of solar maxima and minima, respectively. The horizontal line marks the position of the CZ base.}
    \label{fig:extent}
\end{figure}

We find that both the position, and the width of the tachocline change with time. The results are statistically significant only for some epochs --- for instance, the width changed considerably during solar cycle~23, but less so during cycle~24. The opposite is true for the change in the position of the tachocline --- the results for cycle~23 are confusing, but there is a clear signature of change in cycle~24. During cycle~24, the tachocline was deeper in the radiative zone just before the  maximum, then moved closer to the convection zone in the middle of the descending phase, and started going deeper again. 

Given that both  the position and the width of the tachocline vary with time, it is instructive to look at the upper and lower boundaries  of the tachocline, i.e., $r_d+w_d$ and $r_d-wd$ as is shown in Fig.~\ref{fig:extent}; recall that this region spans the radius range where the rotation rate changes from $0.269\,\delta\Omega$ to $0.731\,\delta\Omega$.  While this information is available in Fig.~\ref{fig:all}, this representation offers a better visualization of the changes.  If we look at the results derived from the two intermediate-length time series, i.e., the $10\times72$-day and $16\times72$-day long ones, it appears that the tachocline becomes narrower around the time of solar maximum. The extent increases after the maximum and starts decreasing again.  Since the maximum of solar cycle~25 is imminent, another a few more  years of helioseismic data should be able to show whether this is indeed the case, or whether noise in the results is fooling us. 

Our results reinforce the  need to continue helioseismic monitoring of the Sun to understand solar activity and its evolution.

\section*{Acknowledgements}

This work is supported by NASA grant 80NSSC23K0563 to SB.
and NASA grants 80NSSC22K0516 and  NNH18ZDA001N-DRIVE to SGK.
This work utilizes data from the National Solar Observatory Integrated Synoptic Program, which is operated by the Association of Universities for Research in Astronomy, under a cooperative agreement with the National Science Foundation and with additional financial support from the National Oceanic and Atmospheric Administration, the National Aeronautics and Space Administration, and the United States Air Force. The GONG network of instruments is hosted by the Big Bear Solar Observatory, High Altitude Observatory, Learmonth Solar Observatory, Udaipur Solar Observatory, Instituto de Astrof\'{\i}sica de Canarias, and Cerro Tololo Interamerican Observatory.
This work also uses provided by the SOHO/MDI consortium. SOHO is a project of international cooperation between ESA and NASA. We also use data from the Helioseismic and Magnetic Imager on board the Solar Dynamics Observatory. HMI data are courtesy of NASA/SDO and the AIA, EVE, and HMI science teams.

\facilities{GONG, MDI, HMI, Royal Observatory of Belgium, Brussels}

\bibliography{main}{}
\bibliographystyle{aasjournal}

\appendix

\section{Fitting methods} \label{app:fitting}

We considered two different forward modeling techniques to derive the three parameters that characterize our representation of the tachocline. The first is simply a search on a pre-computed grid of models, inspired by techniques used in asteroseismology \citep[e.g.,][]{stello, gai, victor}, while the second is a simulated annealing minimization. 

\subsection{Grid-based fitting}

\begin{figure*}[]
    \centering
    \includegraphics[width=\textwidth]{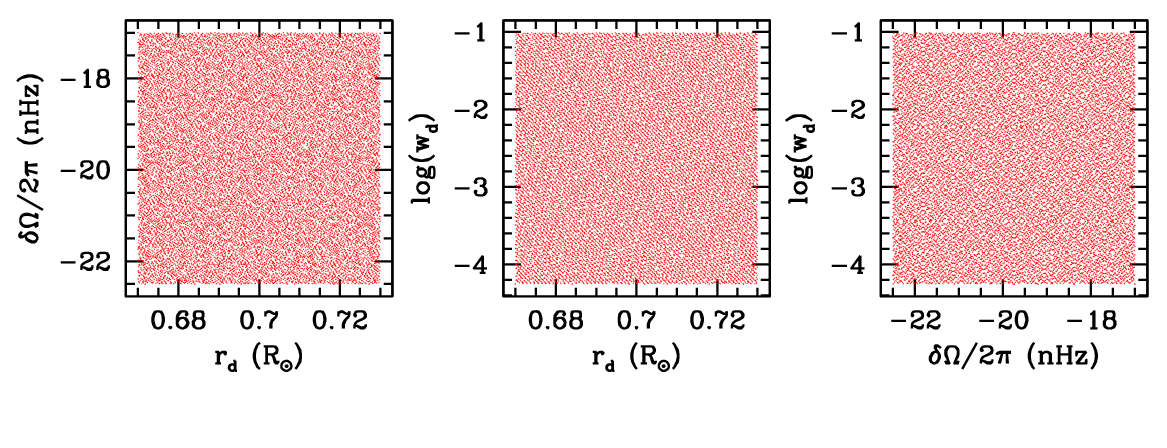}
    \caption{ The distribution of tachocline parameters in the grid of 15000 models. }
    \label{fig:sobol}
\end{figure*}

For the grid-based fitting method, we calculate the $c_3$ splitting coefficients of models with different combinations of $r_d$, 
$\Delta\Omega$ and $w_d$. The parameters are distributed in a 3D Sobol sequence \citep{sobol} of length 15000 to ensure uniformity of coverage of the three-dimensional parameter space. We use the code  from \citet{sobolkuo1, sobolkuo2}\footnote{https://web.maths.unsw.edu.au/~fkuo/sobol/} to determine the Sobol numbers.  The coverage of the parameters is shown in Fig.~\ref{fig:sobol}; note that since the width of the tachocline is small \citep{abc, antiabasu2011}, we choose a distribution of $\log(w_d)$ rather than $w_d$.

We follow the method proposed for \citet{gai}, namely for each model, we calculate the $\chi^2$ between the observed and model splittings. The $\chi^2$, assuming a Gaussian distribution of errors, can be converted to a likelihood for each model as follows:
\begin{equation}
    {\mathcal L}_i=\frac{\exp(-{\chi^2/2})}{\sum_{i=1}^N \exp(-{\chi^2/2})},
    \label{eq:like}
\end{equation}
where $N$ is the total number of models. The parameters of the tachocline are determined by averaging the parameters of the models  with likelihood ${\mathcal L}\ge 0.90\,{\mathcal L}_{\rm max}$, where
${\mathcal L}_{\rm max}$ is the maximum  likelihood obtained. 
 The uncertainties are determined using a
traditional bootstrapping method of simulating many realizations of the
observations, fit each one of them in exactly the same manner as the original
data and use the spread as a measure of uncertainty.

\subsection{Simulated annealing}

We use simulated annealing  \citep[hereafter SA;][]{anneal1,anneal2} to minimize the $\chi^2$ between computed and observed values of $c_3$. This algorithm uses randomly generated values of the parameters, and we assume that the parameters have Gaussian priors with the mean and width of the Gaussian determined from existing inversions of rotational splittings. These inversions have clearly shown the presence of a tachocline, but do not fully resolve it, since regularization is achieved via some form of smoothing.  
Given that there is a chance that the solution becomes trapped in a local minimum, we make 80 different attempts using different sequences of randomly selected initial guesses in order to derive a global $\chi^2$ minimum. We can be certain that the algorithm reached a global minimum by inspecting the likelihood weighted distribution of all the parameters for all iterations, where the likelihood is defined as $\exp(-\chi^2/2)$. Indeed, when  a global minimum has been reached the distribution is single peaked, otherwise it will have multiple peaks. We found that $\delta\Omega$ is the first parameter to converge, followed by $r_d$; $w_d$ takes the longest to converge. As with the grid based case, we determine the uncertainties using a bootstrapping method. Earlier works \citep{abc, antiabasu2011, basuantia2019} used the same approach. 

\begin{figure*}[]
    \centering
    \includegraphics[width=0.8\textwidth]{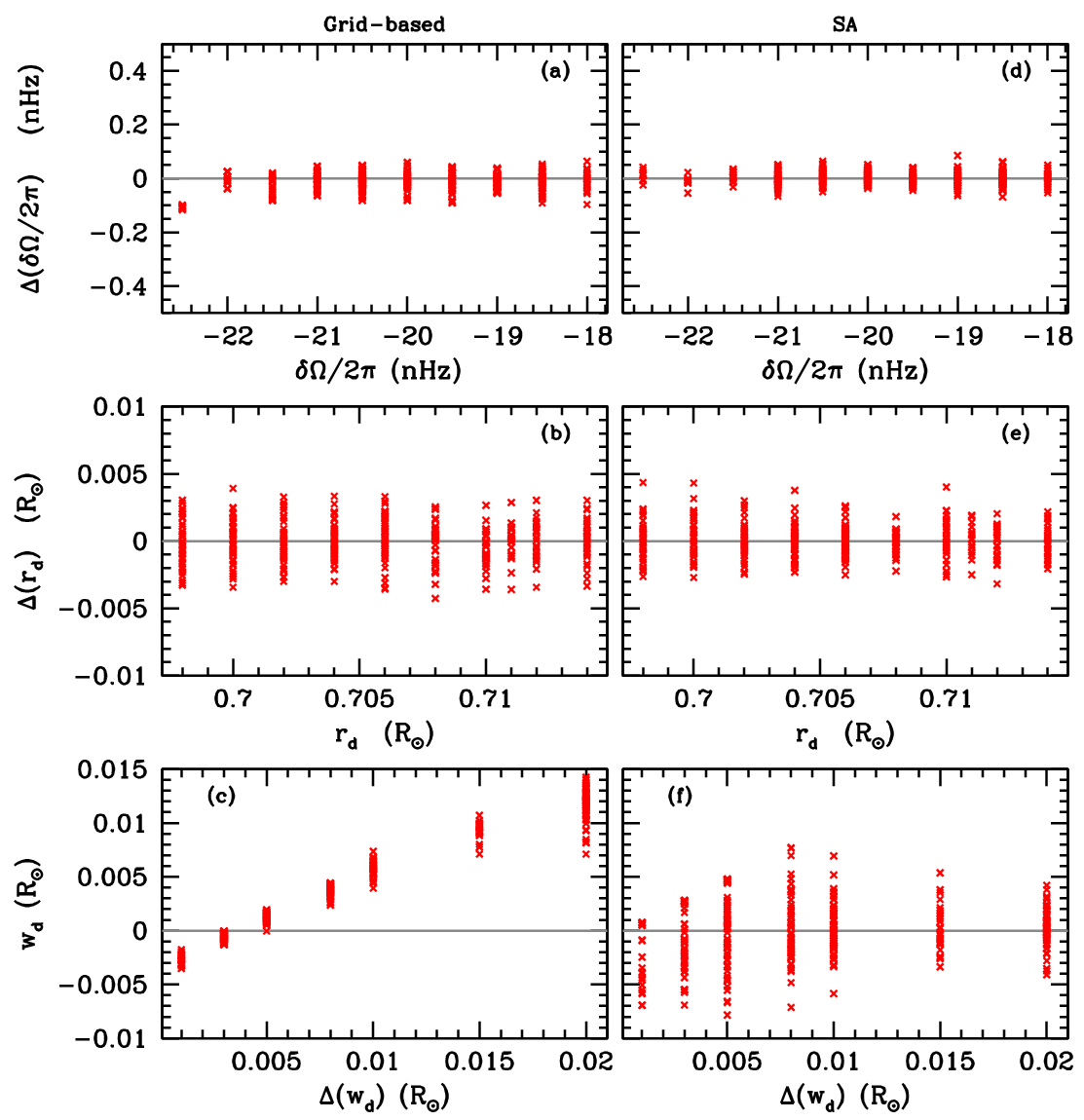}
    \caption{ The difference between the exact and estimated values of tachocline properties plotted as a function of the exact value of the parameters. The differences are in the sense (Exact $-$ Estimate). The left-hand panels (a, b, and c) show results for the grid-based method, the right-hand panels (d, e, and f) show results from simulated annealing.}
    \label{fig:test}
\end{figure*}
\begin{figure*}[]
    \centering
    \includegraphics[width=0.95\textwidth]{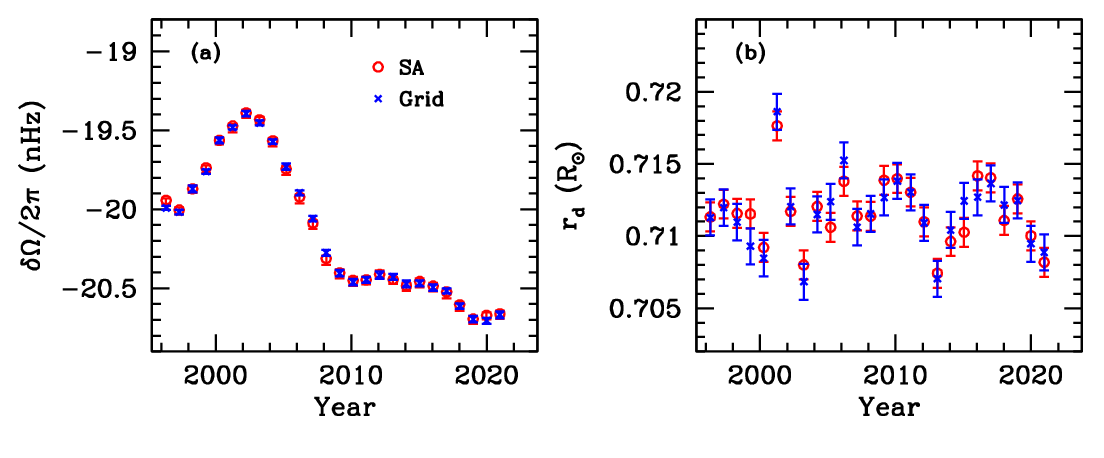}
    \caption{Estimates of $r_d$ and $\Delta\Omega$ for different sets of GONG splittings obtained with $10\times72$-day long time series.}
    \label{fig:comp}
\end{figure*}

\subsection{Comparison}

Test results obtained with the grid and SA methods are presented in Fig.~\ref{fig:test}, where we show results obtained for models with known tachocline properties; we included uncertainties that correspond to those for a $10\times72$-day long HMI data set. 

The figure shows clearly that both the grid-based and simulated annealing methods perform rather well for the jump, $\Delta\Omega$, and the position, $r_d$, parameters. The resulting spread is consistent with the uncertainties obtained by bootstrapping. The one notable exception are the grid-based results for models with the largest jump amplitude, $|\delta\Omega|$; 
this can be attributed to the fact that these models fall at the edge of our Sobol distribution, and hence the resulting truncation causes a systematic error.
Results obtained for the width parameter, $w_d$, are not as good. The SA results show a systematic error at very low values, i.e., the thickness is generally overestimated in those cases, the results are good for larger $w_d$. In contrast, the grid-based method does very poorly overall: it overestimates the width at the low end, and underestimates it at the high end.  

Since one of the aims of this work is to determine possible changes in width, $w_d$, we cannot use a method that has width-dependent systematic errors. Hence, we show only results obtained with the SA method. As shown in Fig.~\ref{fig:comp}, the values of $r_d$ and $\delta\Omega$ obtained by the two methods are commensurate with each other, suggesting that we could cut down on computational time by determining $r_d$ and $\delta\Omega$ using the grid-based method but use the SA method, keeping $r_d$ and $\delta\Omega$ fixed, to derive the width, $w_d$, however, we chose not to do so.

\newpage

\section{Tables with results} \label{app:tabs}

This Appendix lists all our results. Note that once the analyzed time series becomes long enough, the uncertainties in the tachocline parameters become symmetric. The start date of each set has the format YYYYMMDD.

\begin{table*}[h]
  \centering
  \caption{Jump of the tachocline from $72$-day GONG SGK sets}
  \scriptsize
\begin{tabular}{lccccccc}
\toprule
Start Date &$\delta\Omega/2pi$ && Start Date &$\delta\Omega/2\pi$ && Start Date &$\delta\Omega/2\pi$  \\
 & (nHz) && & (nHz) && & (nHz) \\
\hline
\\
19950507 &   $-19.8566^{+0.0182}_{-0.0516}$ &&   20041022 &   $-19.6214^{ +0.0241}_{-0.0357}$ &&  20140409 &   $-20.4649^{ +0.0197}_{-0.0586}$ \\
19950718 &   $-19.6371^{+0.0236}_{-0.0438}$ &&   20050102 &   $-19.5886^{ +0.0383}_{-0.0245}$ &&  20140620 &   $-20.7384^{ +0.0392}_{-0.0217}$ \\
19950928 &   $-20.0472^{+0.0360}_{-0.0260}$ &&   20050315 &   $-19.7492^{ +0.0257}_{-0.0410}$ &&  20140831 &   $-20.4938^{ +0.0237}_{-0.0466}$ \\
19951209 &   $-20.0563^{+0.0248}_{-0.0521}$ &&   20050526 &   $-19.8756^{ +0.0170}_{-0.0373}$ &&  20141111 &   $-20.5675^{ +0.0254}_{-0.0513}$ \\
19960219 &   $-20.0845^{+0.0395}_{-0.0354}$ &&   20050806 &   $-19.7622^{ +0.0244}_{-0.0459}$ &&  20150122 &   $-20.5999^{ +0.0265}_{-0.0749}$ \\
\end{tabular}
\label{tab:01e}
\end{table*}

\newpage
\begin{table*}[h]       
\centering       
\caption{Jump of the tachocline from 72-day MDI SGK sets}       
\scriptsize       
\begin{tabular}{lccccccc}       
\toprule       
Start Date & $\delta\Omega/2\pi$ && Start Date & $\delta\Omega/2\pi$ && Start Date & $\delta\Omega/2\pi$ \\       
 & (nHz) && & (nHz) && & (nHz) \\       
\hline       
\\                                                  
19960501  &   $-20.2503^{+0.0015}_{-0.0062} $ &&   20010827  &  $ -19.5587^{+0.0021}_{-0.0050} $ &&   20060801  &  $ -20.0478^{+0.0042}_{-0.0023} $\\ 
19960712  &   $-20.1691^{+0.0017}_{-0.0059} $ &&   20011107  &  $ -19.3168^{+0.0020}_{-0.0063} $ &&   20061012  &  $ -20.1769^{+0.0019}_{-0.0043} $\\ 
19960922  &   $-20.1874^{+0.0028}_{-0.0042} $ &&   20020118  &  $ -19.4403^{+0.0028}_{-0.0029} $ &&   20061223  &  $ -20.1697^{+0.0035}_{-0.0050} $\\ 
19961203  &   $-20.0857^{+0.0019}_{-0.0043} $ &&   20020331  &  $ -19.5229^{+0.0020}_{-0.0058} $ &&   20070305  &  $ -20.2580^{+0.0033}_{-0.0049} $\\ 
19970213  &   $-20.1462^{+0.0027}_{-0.0069} $ &&   20020611  &  $ -19.5605^{+0.0019}_{-0.0043} $ &&   20070516  &  $ -20.4187^{+0.0026}_{-0.0048} $\\
\end{tabular} 
\label{tab:mdi01e} 
\end{table*} 

\begin{table*}[h]
\caption{Jump of the tachocline from 72-day HMI SGK sets}              
\scriptsize              
\begin{tabular}{lccccccc}                   
\toprule              
Start Date & $\delta\Omega/2\pi$ && Start Date & $\delta\Omega/2\pi$ && Start Date & $\delta\Omega/2\pi$ \\              
 & (nHz) && & (nHz) & (nHz) \\              
\hline              
\\         
20100430 &  $-20.4498^{+0.0028}_{-0.0039}$ && 20141111 &  $-20.7100^{+0.0039}_{-0.0046}$ && 20190525 &  $-20.7434^{+0.0025}_{-0.0041}$ \\  
20100711 &  $-20.5844^{+0.0028}_{-0.0052}$ && 20150122 &  $-20.4783^{+0.0024}_{-0.0038}$ && 20190805 &  $-20.8289^{+0.0022}_{-0.0068}$ \\  
20100921 &  $-20.6022^{+0.0027}_{-0.0052}$ && 20150404 &  $-20.4555^{+0.0015}_{-0.0055}$ && 20191016 &  $-20.7255^{+0.0026}_{-0.0012}$ \\  
20101202 &  $-20.4409^{+0.0021}_{-0.0051}$ && 20150615 &  $-20.5713^{+0.0024}_{-0.0017}$ && 20191227 &  $-20.7754^{+0.0027}_{-0.0034}$ \\  
20110212 &  $-20.4206^{+0.0027}_{-0.0041}$ && 20150826 &  $-20.4761^{+0.0018}_{-0.0034}$ && 20200308 &  $-20.7407^{+0.0018}_{-0.0054}$ \\
\end{tabular}        
\label{tab:hmi01e}        
\end{table*}

\begin{table*}[h]  
  \centering  
  \caption{Tachocline properties from $5\times72$-day SGK sets}  
  \scriptsize  
\begin{tabular}{lcccccccc}  
\toprule
Start Date &&   $\delta\Omega/2\pi$    &            $r_d$     &            $w_d$      &&            $\delta\Omega/2\pi$  &              $r_d$ &                $w_d$  \\ 
           &&   (nHz)                  &             (R$_\odot$) &       (R$_\odot$) &&  (nHz) & (R$_\odot$) &       (R$_\odot$) \\
\hline           
 & &\multicolumn{3}{c}{Asymmetric Fits}& &\multicolumn{3}{c}{Symmetric Fits}\\ 
\hline 
&&\multicolumn{6}{c}{GONG} \\
\hline
19950507  &&   $-20.0690^{+0.0148}_{-0.0245} $ &   $0.7154^{+0.0007}_{-0.0017}$ &    $0.0037^{+0.0076}_{-0.0030} $ &&   $-20.0654^{+0.0138}_{-0.0299} $&    $0.7170^{+0.0007}_{-0.0015}$ &    $0.0050^{+0.0074}_{-0.0030}$  \\ 
19960501  &&   $-20.0673^{+0.0158}_{-0.0264} $ &   $0.7158^{+0.0007}_{-0.0017}$ &    $0.0036^{+0.0076}_{-0.0029} $ &&   $-20.0654^{+0.0138}_{-0.0299} $&    $0.7170^{+0.0007}_{-0.0015}$ &    $0.0080^{+0.0074}_{-0.0030}$  \\ 
19970426  &&   $-19.9135^{+0.0218}_{-0.0244} $ &   $0.7133^{+0.0012}_{-0.0019}$ &    $0.0116^{+0.0036}_{-0.0022} $ &&   $-19.9322^{+0.0217}_{-0.0241} $&    $0.7145^{+0.0013}_{-0.0019}$ &    $0.0127^{+0.0036}_{-0.0023}$  \\ 
19980421  &&   $-19.8334^{+0.0161}_{-0.0272} $ &   $0.7119^{+0.0011}_{-0.0018}$ &    $0.0084^{+0.0056}_{-0.0015} $ &&   $-19.8347^{+0.0196}_{-0.0244} $&    $0.7131^{+0.0011}_{-0.0020}$ &    $0.0088^{+0.0054}_{-0.0016}$  \\
\end{tabular}  
\label{tab:05a}  
\end{table*}  

\newpage
\begin{table*}[h]   
  \centering   
  \caption{Tachocline from $5\times72$-day official pipeline sets}   
  \scriptsize   
\begin{tabular}{lcccclccc}   
\toprule 
Start Date &   $\delta\Omega/2\pi$    &            $r_d$     &            $w_d$      &&   Start Date &         $\delta\Omega/2\pi$  &              $r_d$ &                $w_d$  \\  
           &  (nHz)                  &             (R$_\odot$) &       (R$_\odot$) &&  & (nHz) & (R$_\odot$) &       (R$_\odot$) \\  
\hline 
\multicolumn{4}{c}{Asymmetric Fits}&&\multicolumn{4}{c}{Symmetric Fits}\\  
\hline 
\multicolumn{9}{c}{MDI} \\ 
\hline    
19960501 &   $-19.7402^{+0.0154}_{-0.0467}$ &    $0.7130^{+0.0011}_{-0.0021}$  &   $0.0090^{+0.0056}_{-0.0017}$  && 19960501 &    $-19.7107^{+0.0142}_{-0.0565}$  &   $0.7190^{+0.0011}_{-0.0021}$  &   $0.0074^{+0.0054}_{-0.0014}$  \\ 
19970426 &   $-19.5804^{+0.0195}_{-0.0454}$ &    $0.7146^{+0.0014}_{-0.0022}$  &   $0.0138^{+0.0034}_{-0.0024}$  && 19960712 &    $-19.7026^{+0.0210}_{-0.0464}$  &   $0.7197^{+0.0012}_{-0.0023}$  &   $0.0112^{+0.0039}_{-0.0017}$  \\ 
19980421 &   $-19.2731^{+0.0140}_{-0.0451}$ &    $0.7104^{+0.0008}_{-0.0019}$  &   $0.0060^{+0.0080}_{-0.0032}$  && 19970426 &    $-19.5795^{+0.0152}_{-0.0427}$  &   $0.7144^{+0.0014}_{-0.0021}$  &   $0.0129^{+0.0024}_{-0.0021}$  \\ 
19990416 &   $-19.2008^{+0.0216}_{-0.0390}$ &    $0.7161^{+0.0012}_{-0.0023}$  &   $0.0101^{+0.0051}_{-0.0019}$  && 19970707 &    $-19.5339^{+0.0214}_{-0.0426}$  &   $0.7163^{+0.0013}_{-0.0022}$  &   $0.0130^{+0.0029}_{-0.0018}$  \\
\end{tabular}   
\label{tab:05b}  
\end{table*}
\newpage

\begin{table*}[h]   
  \centering   
  \caption{Tachocline properties from  $8\times72$-day SGK sets}   
  \scriptsize   
\begin{tabular}{l |ccc  |ccc} 
\toprule 
\hline 
Start Date & $\delta\Omega/2\pi$    &            $r_d$     &            $w_d$      &  $\delta\Omega/2\pi$    &            $r_d$     &            $w_d$      \\ 
           &   (nHz)               &           (R$_\odot$) &       (R$_\odot$)     &  (nHz)                  & (R$_\odot$)          &       (R$_\odot$)     \\ 
\hline 
&\multicolumn{3}{c}{GONG}&\multicolumn{3}{c}{MDI}\\
19950718  &   $-19.9715\pm 0.0117$  &     $ 0.7090\pm 0.0012$ &    $ 0.0098\pm 0.0017$  &                             &                      &                           \\ 
19960501  &   $-20.0201\pm 0.0182$  &     $ 0.7131\pm 0.0011$ &    $ 0.0114\pm 0.0016$  &    $-20.0907\pm    0.0194$ &   $ 0.7144\pm 0.0011$ &     $ 0.0128\pm 0.0016$  \\ 
19970213  &   $-19.9300\pm 010248$  &     $ 0.7098\pm 0.0011$ &    $ 0.0148\pm 0.0015$  &    $-19.9548\pm    0.0138$ &   $ 0.7083\pm 0.0012$ &     $ 0.0149\pm 0.0016$  \\ 
19971128  &   $-19.8463\pm 0.0154$  &     $ 0.7119\pm 0.0012$ &    $ 0.0106\pm 0.0016$  &    $-19.8492\pm    0.0196$ &   $ 0.7098\pm 0.0011$ &     $ 0.0063\pm 0.0016$  \\ 
19980912  &   $-19.7477\pm 0.0113$  &     $ 0.7101\pm 0.0011$ &    $ 0.0090\pm 0.0013$  &    $-19.6535\pm    0.0163$ &   $ 0.7089\pm 0.0011$ &     $ 0.0108\pm 0.0016$  \\
\end{tabular}   
\label{tab:08}   
\end{table*}    

\newpage

\begin{table*}[h]
  \centering
  \caption{Tachocline properties from  $10\times72$-day SGK sets}
  \scriptsize
\begin{tabular}{l | ccc | ccc}
\toprule
\hline
Start Date & $\delta\Omega/2\pi$    &            $r_d$     &            $w_d$      &  $\delta\Omega/2\pi$    &            $r_d$     &            $w_d$      \\
           &   (nHz)               &           (R$_\odot$) &       (R$_\odot$)     &  (nHz)                  & (R$_\odot$)          &       (R$_\odot$)     \\
\hline
&\multicolumn{3}{c}{GONG}& \multicolumn{3}{c}{MDI}\\
19950507  &  $-19.9427\pm 0.0118$  &    $0.7113\pm 0.0010$   &   $0.0094\pm 0.0012$ \\
19960501  &  $-20.0048\pm 0.0154$ &     $0.7122\pm 0.0012$  &    $0.0122\pm 0.0014$  &     $-20.0204\pm 0.0150$ &  $ 0.7105\pm 0.0009$ &  $ 0.0147\pm 0.0014$  \\
19970426  &  $-19.8701\pm 0.0214$ &     $0.7116\pm 0.0010$  &    $0.0093\pm 0.0013$  &     $-19.9117\pm 0.0194$ &  $ 0.7097\pm 0.0012$ &  $ 0.0080\pm 0.0014$  \\
19980421  &  $-19.7365\pm 0.0147$ &     $0.7115\pm 0.0010$  &    $0.0067\pm 0.0012$  &     $-19.6516\pm 0.0121$ &  $ 0.7108\pm 0.0010$ &  $ 0.0066\pm 0.0013$  \\
19990416  &  $-19.5627\pm 0.0185$ &     $0.7092\pm 0.0010$  &    $0.0077\pm 0.0011$  &     $-19.5359\pm 0.0179$ &  $ 0.7090\pm 0.0013$ &  $ 0.0074\pm 0.0013$  \\
\end{tabular}   
\label{tab:10}   
\end{table*}

\newpage

\begin{table*}[h]    
  \centering    
  \caption{Tachocline properties from  $16\times72$-day SGK sets}    
  \scriptsize    
\begin{tabular}{l|ccc |ccc}  
\toprule  
\hline  
Start Date & $\delta\Omega/2\pi$    &            $r_d$     &            $w_d$      &  $\delta\Omega/2\pi$    &            $r_d$     &            $w_d$      \\  
           &   (nHz)               &           (R$_\odot$) &       (R$_\odot$)     &  (nHz)                  & (R$_\odot$)          &       (R$_\odot$)     \\  
\hline  
&\multicolumn{3}{c}{GONG}& \multicolumn{3}{c}{MDI}\\ 
19960501  &   $ -19.9262\pm  0.0107$  &   $ 0.7137\pm  0.0008$  &   $ 0.0076\pm  0.0011$ &     $-19.9428\pm  0.0110$  &    $ 0.7127\pm  0.0008$   & $    0.0092\pm  0.0011$     \\ 
19971128  &   $ -19.7035\pm  0.0115$  &   $ 0.7110\pm  0.0009$  &   $ 0.0086\pm  0.0010$ &     $-19.6422\pm  0.0118$  &    $ 0.7109\pm  0.0009$   & $    0.0094\pm  0.0012$     \\ 
19990627  &   $ -19.4843\pm  0.0171$  &   $ 0.7139\pm  0.0010$  &   $ 0.0078\pm  0.0010$ &     $-19.4708\pm  0.0118$  &    $ 0.7143\pm  0.0010$   & $    0.0116\pm  0.0011$     \\ 
20010123  &   $ -19.4542\pm  0.0153$  &   $ 0.7123\pm  0.0008$  &   $ 0.0143\pm  0.0010$ &     $-19.4566\pm  0.0125$  &    $ 0.7123\pm  0.0009$   & $    0.0142\pm  0.0013$     \\ 
20020822  &   $ -19.5802\pm  0.0126$  &   $ 0.7103\pm  0.0008$  &   $ 0.0166\pm  0.0011$ &     $-19.6231\pm  0.0112$  &    $ 0.7111\pm  0.0008$   & $    0.0158\pm  0.0013$     \\
\end{tabular}    
\label{tab:16}    
\end{table*}     

 \begin{table*}[h]     
  \centering     
  \caption{Tachocline properties from  $32\times72$-day SGK sets}     
  \scriptsize     
\begin{tabular}{l|ccc |ccc}   
\toprule   
\hline   
Start Date & $\delta\Omega/2\pi$    &            $r_d$     &            $w_d$      &  $\delta\Omega/2\pi$    &            $r_d$     &            $w_d$      \\   
           &   (nHz)               &           (R$_\odot$) &       (R$_\odot$)     &  (nHz)                  & (R$_\odot$)          &       (R$_\odot$)     \\   
\hline   
&\multicolumn{3}{c}{GONG}& \multicolumn{3}{c}{MDI}\\  
19960501   &  $-19.7033\pm 0.0157$  &   $ 0.7134\pm 0.0005$  &   $ 0.0081\pm 0.0007$  &            $ -19.6745\pm 0.0104$  &  $  0.7132\pm 0.0006$  &   $   0.0106\pm 0.0008$  \\ 
19990627   &  $-19.5416\pm 0.0163$  &   $ 0.7121\pm 0.0006$  &   $ 0.0129\pm 0.0007$  &            $ -19.5264\pm 0.0117$  &  $  0.7129\pm 0.0005$  &   $   0.0138\pm 0.0007$  \\ 
20020822   &  $-19.8706\pm 0.0156$  &   $ 0.7119\pm 0.0006$  &   $ 0.0161\pm 0.0008$  &            $ -19.8758\pm 0.0124$  &  $  0.7124\pm 0.0007$  &   $   0.0166\pm 0.0008$  \\ 
20051017   &  $-20.2758\pm 0.0131$  &   $ 0.7143\pm 0.0004$  &   $ 0.0131\pm 0.0006$  &\\ 
20081212   &  $-20.4341\pm 0.0132$  &   $ 0.7120\pm 0.0006$  &   $ 0.0140\pm 0.0009$  &\multicolumn{3}{c}{HMI}\\ 
20100711   &  $-20.4431\pm 0.0116$  &   $ 0.7109\pm 0.0003$  &   $ 0.0124\pm 0.0008$  &            $ -20.4326\pm 0.0144$  &  $   0.711\pm  0.000$6 &   $   0.0120\pm 0.0008$  \\ 
20120207   &  $-20.4788\pm 0.0159$  &   $ 0.7127\pm 0.0006$  &   $ 0.0131\pm 0.0007$  &            $ -20.4788\pm 0.0108$  &  $   0.712\pm  0.000$4 &   $   0.0135\pm 0.0007$  \\ 
20150404   &  $-20.5943\pm 0.0116$  &   $ 0.7118\pm 0.0005$  &   $ 0.0121\pm 0.0008$  &            $ -20.5774\pm 0.0127$  &  $   0.711\pm  0.000$5 &   $   0.0122\pm 0.0008$  \\ 
\toprule   
\end{tabular}     
\label{tab:32}     
\end{table*}

\end{document}